\documentclass{WileyMSP-template}

\usepackage{xcolor}
\usepackage{multirow}
\usepackage{array}
\usepackage{amsmath}
\usepackage{graphicx}
\usepackage{threeparttable}
\usepackage[normalem]{ulem}
\usepackage{makecell}

\begin{document}


\title{Interdigitated Terahertz Metamaterial Sensors: Design with the Dielectric Perturbation Theory}

\maketitle


\author{Lei Cao}
\author{Fanqi Meng*}
\author{Esra Özdemir}
\author{Yannik Loth}
\author{Merle Richter}
\author{Anna Katharina Wigger}
\author{Maira Pérez Sosa}
\author{Alaa Jabbar Jumaah}
\author{Shihab Al-Daffaie}
\author{Peter Haring Bolívar}
\author{Hartmut G. Roskos*}



\begin{affiliations}
Dr. Lei Cao\\
Address\\State Key Laboratory of Advanced Electromagnetic Technology, Huazhong University of Science and Technology, Wuhan 430074, China\\
Address\\Physikalisches Institut, Johann Wolfgang Goethe-Universit\"{a}t, Frankfurt am Main, Germany\\

Dr. Fanqi Meng\\
Address\\Physikalisches Institut, Johann Wolfgang Goethe-Universit\"{a}t, Frankfurt am Main, Germany\\
Email Address:fmeng@physik.uni-frankfurt.de

Esra Özdemir\\
Address\\Physikalisches Institut, Johann Wolfgang Goethe-Universit\"{a}t, Frankfurt am Main, Germany\\

Yannik Loth\\
Address\\University of Siegen, Institute for High Frequency and Quantum Electronics, Siegen, Germany\\

Merle Richter\\
Address\\University of Siegen, Institute for High Frequency and Quantum Electronics, Siegen, Germany\\

Dr.~Anna Katharina Wigger\\
Address\\University of Siegen, Institute for High Frequency and Quantum Electronics, Siegen, Germany\\

Maira Pérez Sosa\\
Address\\Department of Electrical Engineering, Eindhoven University of Technology, 5612 AE Eindhoven, Netherlands\\

Dr. Alaa Jabbar Jumaah\\
Address\\Department of Electrical Engineering, Eindhoven University of Technology, 5612 AE Eindhoven, Netherlands

Dr. Shihab Al-Daffaie\\
Address\\Department of Electrical Engineering, Eindhoven University of Technology, 5612 AE Eindhoven, Netherlands

Prof. Peter Haring Bolívar\\
Address\\University of Siegen, Institute for High Frequency and Quantum Electronics, Siegen, Germany

Prof. Hartmut G. Roskos\\
Address\\Physikalisches Institut, Johann Wolfgang Goethe-Universit\"{a}t, Frankfurt am Main, Germany\\
Email Address:roskos@physik.uni-frankfurt.de

\end{affiliations}



\begin{abstract}
Designing terahertz sensors with high sensitivity to detect nanoscale thin films and single biomolecule presents a significant challenge, and addressing these obstacles is crucial for unlocking their full potential in scientific research and advanced applications. This work presents a strategy for the design optimization of metamaterial sensors employed in the detection of small amounts of dielectric materials. The sensors usually utilize the shift of the resonance frequency as an indicator of the presence of the analyte. The amount of shifting depends on intrinsic properties (electric field distribution, quality factor, and mode volume) of the bare cavity, as well as the overlap volume of its high-electric-field zone(s) and the analyte.
Guided by the simplified dielectric perturbation theory, interdigitated electric split-ring resonators (ID-eSRR) are devised to significantly enhance the detection sensitivity for thin-film analytes compared to eSRRs without interdigitated fingers in the SRR gap region. The fingers of the ID-eSRR metamaterial sensor redistribute the electric field, creating strongly localized field enhancements that substantially boost the interaction with the analyte. Additionally, the periodic change of the orientation of the inherent anti-phase electric field in the interdigitated structure reduces radiation loss, leading to a higher Q-factor.
Experiments with e-beam-fabricated ID-eSRR sensors operating at around 300~GHz demonstrate a remarkable frequency shift of 33.5~GHz upon deposition of a $\mathrm{SiO}_2$ layer with a thickness of 150~nm as an analyte simulant. The figure of merit (FOM) improves by over 50 times compared to structures without interdigitated fingers. This rational design option opens a promising avenue for highly sensitive detection of thin films and trace biomolecules.
\end{abstract}

\section{Introduction}
Metamaterials (MMs) are artificial periodic structures with sub-wavelength features. Their response to electromagnetic radiation can be controlled by their structural design and the materials they are composed of \cite{Withayachumnankul09}. The strong local electric field enhancement by the subwavelength mode concentration makes MMs ideal candidates for sensing analytes. The exploration of applications of MMs for sensing has been flourishing in the last two decades. We concentrate here on planar metallic MMs used at terahertz (THz) frequencies \cite{Debus07, Al-Naib08, Singh14, Weisenstein20, Park17, Meng22}. Frequently used unit-cell structures of metallic MMs consist of square or circular split-ring resonators (SRRs) \cite{Withayachumnankul12, Islam17}, asymmetric double-split-ring resonators (aDSRRs) \cite{Debus07, Al-Naib08, Singh14}, electric split-ring resonators (eSRRs) \cite{Park17, Jun22}, I-shaped structures \cite{Gupta20}, cross-shaped structures \cite{Hu16},\\ 
labyrinthine structures \cite{Jauregui18, Saadeldin19}, and coupled structures with two or more resonators \cite{Driscoll07, Al-Naib17, Jin18, Ma20author, Sun22, Meng23}.
Although THz sensors based on metallic MMs have shown advantages in terms of contact- and label-free detection \cite{Tang20, Gezimati23}, they still have disadvantages in terms of sensitivity and detection limits compared to existing mature and standardized biochemical diagnostic methods such as enzymatic immunoassays \cite{Weisenstein21, Richter22}. Therefore, further improvements in the sensitivity of THz sensors are a crucial task in order to make this sensing method competitive for practical applications.

It is widely accepted that optimized MM sensors should exhibit a strong local electric field enhancement at the location of the analyte, a high quality factor (Q-factor) to ensure a narrow spectral resonance for reliable identification of a resonance shift, and a small mode volume—an aspect recently recognized as a key parameter—in order to achieve a large overlap of the electric field with the analyte \cite{Gupta20}. However, a systematic and practical optimization strategy that coherently considers all the parameters for the rational design of MM sensors has been missing until now.

Another shortcoming in this field of sensor research relates to the performance indicators usually employed. For the assessment of the relative performance of MM sensors, two parameters find application. The first is the \textit{refractive index sensitivity}, often denoted by the letter \textit{S}, which is calculated as the resonance frequency shift resulting from a change in the refractive index of the analyte by unity (and hence is measured in units of 'GHz/RIU', RIU standing for 'refractive index unit'). The second indicator is the \textit{figure of merit} (FOM), calculated as the ratio of \textit{S} to the full width at half maximum (FWHM) of the resonance line in the power spectrum (and hence is measured in units of RIU$^{-1}$) \cite{Askari21author}. The problem with these parameters is that they only allow the comparison of sensitivities among various sensors if the analyte is applied at the same location and in the same amount, which means thin films should be applied with the same layer thickness. However, this is often disregarded in the literature, as will be shown later. 

Here, we address these two issues. We employ dielectric perturbation theory to guide the rational design of planar MM sensors. Based on this rationale, we propose the interdigitated eSRR (ID-eSRR) with mini-gaps as an optimized structure to detect thin films. The proposed ID-eSRR significantly enhances the detection sensitivity and increases the Q-factor compared to the eSRR without fingers by a large amount. We experimentally verify that the FOM of the designed ID-eSRR, when sensing an analyte in the form of a thin film, improves by more than fifty times compared to the structure without interdigitated fingers.

It also becomes apparent that it is often unconvincing to compare sensors using only those two conventional performance parameters ($S$ and FOM). One should design different sensors for different types of analytes and employ different performance parameters. For analytes brought onto the sensor in the form of continuous thin films, we suggest a third performance indicator obtained by normalizing the FOM by the thickness of the film. We term this indicator a TN-FOM (abbreviation for 'thickness-normalized FOM').

\section{A strategy for the rational design of MM sensors}
The property used for sensing is the frequency shift of one or several resonant modes due to the dielectric load of the analyte. To increase sensitivity, it is necessary to locally enhance the electric field and compact the mode volume at the respective resonance frequencies \cite{Gupta20}. Dielectric perturbation theory is widely employed in the microwave community to predict the resonance frequency shift of cavities \cite{Pozar11, Cao22}. We extend this theory to the THz frequency range and propose a strategy to predict the resonance frequency shift of MM sensors.

In general, the relative change in resonance frequency caused by a non-magnetic analyte in a MM resonator can be calculated analytically:
\begin{equation}
    \frac{\Delta f}{f_0} = -\frac{\Delta W}{W_0} = \frac{-\int_{\Delta V} \Delta \epsilon |\vec{E}_0|^2 dV }{\int_{V} \epsilon |\vec{E}_0|^2 dV +\int_{V}  \mu |\vec{H}_0|^2 dV }  \ \,.
\label{equ:perturbation}
\end{equation}
Here, $W_0$ represents the energy (electric energy $W_e$ and magnetic energy $W_m$) stored in the cavity without the analyte, and $\Delta W$ is the change in that energy due to the presence of the analyte. $\vec{E}_0$ and $\vec{H}_0$ denote the electric and magnetic field distributions, and $\mu$ and $\epsilon$ represent the permeability and permittivity of the filling medium (or media) in the resonator (effective cavity volume $V$), all before the introduction of the analyte. $\Delta \epsilon$ indicates the relative change in permittivity associated with the analyte, which occupies the volume $\Delta V$, compared to the original filling medium.
All quantities in Equation~(\ref{equ:perturbation}), including $\vec{E}_0$, $\vec{H}_0$, $\mu$, $\epsilon$, and $\Delta\epsilon$, are implicitly understood to be spatially dependent functions ($F=F(\vec{r})$). The equation is derived under the assumption that the electric field distribution of the respective resonator mode is only weakly perturbed by the analyte, and the tangential component of the electric field is null at the boundary of the resonator (the boundary in the case of MMs is the delimiter of the unit cell; in our simulations, we extend the vertical boundaries out to 30~$\mu$m above (air) and below (substrate) the surface of the substrate \cite{Gupta20, Cao22}).
According to Equation~(\ref{equ:perturbation}), the frequency shift caused by the analyte depends on the size of the volume $\Delta V$ occupied by the analyte and the spatial dependence of the electric field in the cavity volume (volume integration, giving relevance to the electric field enhancement factor). \\

The stored electric energy and magnetic energy are almost equal at the resonance frequency, as confirmed in the simulations. The denominator on the right side of Equation~(\ref{equ:perturbation}) can hence be approximated as $\int_{V} \epsilon |\vec{E}_0|^2 dV +\int_{V}  \mu |\vec{H}_0|^2 dV = 2\, \int_{V} \epsilon |\vec{E}_0|^2 dV $. At the resonance frequency of the MM structure, the electric field is mainly concentrated in and at the splits (gaps) of the metallic resonators, and especially at the corners of the metal edges. By reducing the width of the gaps, the local electric field is enhanced, and the mode volume is reduced, both of which usually lead to an improvement in sensitivity \cite{Seo09, Bahk17, Adak19}.

\begin{figure}
  \includegraphics[width=0.8\linewidth]{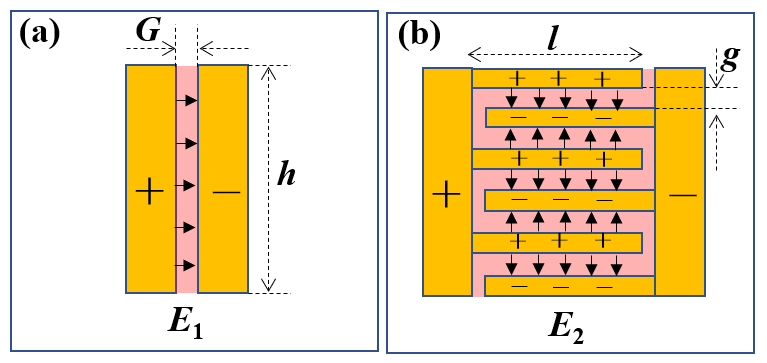}
  \caption{Schematics of two gap structures used in SRRs: (a) One-dimensional slot-like gap and (b) two-dimensional interdigitated gap structure.  
  The areas plotted in pink color represent the regions with high electric field. If $G\,$=$\,g$ and if the same voltages are applied, the electric fields in the gap regions in (a) and (b) are of similar magnitude. 
  }
  \label{fig:design strategy}
\end{figure}

If we assume the analyte to be sufficiently thin and positioned in the region of the maximum electric field, and if the electric field there is assumed to be uniform, Equation~(\ref{equ:perturbation}) can be simplified to
\begin{equation}
    \frac{\Delta f}{f_0} \approx \frac{-\Delta \epsilon \max[|\vec{E}_0|^2] \cdot d \cdot A }{2 \int_{V} \epsilon |\vec{E}_0|^2 dV} =  \frac{-\Delta \epsilon \cdot d \cdot A }{2\epsilon_s V_{eff}} \ \,.
\label{equ:simplified perturbation}
\end{equation}
Here, $d$ is the thickness of the analyte layer and $A$ represents the overlap area between the analyte and the region of high electric field. 
$V_{eff}$ in the denominator on the right side of Equation~(\ref{equ:simplified perturbation}) is the mode volume which quantifies the spatial extent of the electric field of the resonance mode. It is defined by \cite{Gerry04}
\begin{equation}
    V_{eff} = \frac{\int_{V} \epsilon |\vec{E}_0|^2 dV}{\max[\epsilon_s |\vec{E}_0|^2]} 
\label{equ:mdoe volume}
\end{equation}
as the ratio of the total electric energy of the mode and the maximal value of the electric energy density ($\epsilon|\vec{E}_0|^2)$), whose location is usually in the substrate with the dielectric constant $\epsilon_s$. 

The value of $A$ in Equation~(\ref{equ:simplified perturbation}) depends on the field distribution of the MM sensor and the way the analyte is deposited. For our approach, we use the fact that the electric field in many MM structures is concentrated around the metal edges bordering the gaps of the metallic resonators (compare the two cases in \textbf{Figure \ref{fig:design strategy}}, in red: high-field regions in the gap between the metallic edges). Since the electric field distribution is inhomogeneous, a possible way to define the bounds of the high-field region is to specify a delimiter where the electric field amounts to ten percent of the maximal strength of the electric field \cite{Debus07}.
If the analyte is deposited homogeneously as a thin layer covering the whole surface of the MM (which would also be a good assumption in the case of a homogeneous film of dried viral or cellular biomaterial \cite{Park17, Yan19}), then $A$ is the area of overlap between the region of high electric field and the analyte layer. If, on the other hand, the analyte covers only a small area within the high-field region, then $A$ equals the area covered by the analyte. 

According to Equation~(\ref{equ:simplified perturbation}), a small value of $V_{eff}$ helps enhance the frequency shift. The shift is larger if more of the field is concentrated in the analyte. The straightforward way to achieve field concentration is by reducing the width of the gap in the MM structure. This has been explored in the literature \cite{Park17, Seo09, Bahk17, Adak19}. In the experiments of Ref.~\cite{Park17}, for example, an eSRR MM structure was used as a sensor for uniformly dried virus layers. When the gap width was reduced from 3~$\mu$m to 200~nm, the sensitivity increased by a factor of thirteen.

Another important quantity is the linewidth of the resonance (Q-factor). The larger the Q-factor, the easier the frequency shift can be recognized. With the refractive index $n_a$ of the analyte ($n_a \neq 1$), one obtains the sensitivity as $S=|\Delta f/ \Delta n|= |\Delta f/(n_a-1)|$, and using Equation~(\ref{equ:simplified perturbation}), the \text{FOM} value can be approximately expressed as
\begin{equation}
    \text{FOM} = \frac {S}{\text{FWHM}} \approx \frac{\Delta \epsilon \cdot d \cdot A }{2 \, \Delta n} \cdot  \frac{Q}{V_{eff}\cdot \epsilon_s}\ \,.
\label{equ:FOM}
\end{equation}
The first fraction in Equation~(\ref{equ:FOM}) is determined by the parameters of the analyte, and the second by the properties of the MM resonator. The FOM is often used to compare the performance of different sensors. However, this is problematic due to the first fraction. The thickness, the overlap area, and the refractive index of the analyte all change the FOM, making the comparison of different sensors difficult when they are intended for analytes with differing properties. As a universal parameter for the comparison of sensors, the FOM should be \textit{analyte-independent}.

In the microwave sensor community, the sensitivity is not normalized by the refractive index unit ($\Delta n $) but instead by the permittivity unit ($\Delta \epsilon $) \cite{Abdolrazzaghi18, CaoY22}. Then Equation~(\ref{equ:FOM}) can be further simplified because $\Delta \epsilon $ appears in both the numerator and the denominator. The influence of the refractive index (resp. the permittivity) is then eliminated. If one further normalizes by the overlap volume ($d \cdot A$), then a new figure of merit is obtained which contains only the second fraction in Equation~(\ref{equ:FOM}), $Q/(V_{eff}\epsilon_s)$. Such a figure of merit does not depend on the analyte and only reflects the performance of the sensor.
For the universal comparison of sensors, this quantity represents a well-suited performance parameter \cite{Gupta20}. This is especially true and useful for the design of sensors for monitoring minute traces of an analyte (such as biochemicals down to ultra-low molecule concentrations) located only in the high-electric-field region of the sensors.

In the following, we will concentrate on MM sensors, where thin analyte films are homogeneously deposited over the whole surface of the sensor. We assume that the thickness of the film is known. 
For this specific case, we introduce a \textit{thickness-normalized} FOM (TN-FOM), obtained by dividing the conventional FOM by the thickness of the analyte film. To remain consistent with the common practice of the THz MM sensing community, we continue normalizing $S$ and the FOM by the refractive index unit. With Equation~(\ref{equ:FOM}), the TN-FOM is \begin{equation}
    \text{TN-FOM} = \frac {\text{FOM}}{d} = \frac{\Delta \epsilon \cdot A }{2 \Delta n} \cdot  \frac{Q}{V_{eff}\cdot \epsilon_s}\ \,.
\label{equ:TNFOM}
\end{equation}
This quantity is useful for sensor comparison even if analyte materials are different.
As the analyte is deposited as a continuous layer, the overlap area $A$ is determined by the field distribution on the SRR and can hence be considered a property of the sensor itself.  
Having discussed various performance parameters, we now analyze Equation~(\ref{equ:FOM}), respectively, Equation~(\ref{equ:TNFOM}), regarding ways to improve the sensors. One way to achieve this is by the choice of a substrate with a low dielectric constant $\epsilon_s$. This measure has already been explored in the literature \cite{Srivastava19}. We do not pursue this option further in the present paper.

As discussed above, with a smaller gap comes a smaller high-field region $A$ (overlap area), but the ratio $A/V_{eff}$ in the equations significantly increases because $V_{eff}$ is inversely proportional to the maximum electric field, which scales inversely with the width of the gap \cite{Seo09, Bahk17, Adak19}.
Now, the question arises whether an extension of the length $l$ of the gap, for example by giving it a meandering shape as indicated in Figure~\ref{fig:design strategy}, could improve the value of the TN-FOM if one keeps the gap width fixed. At first glance, this should not be the case. If we approximate the overlap area $A$ by the gap area, then an increase in the length of the gap will linearly raise the value of $A$ ($A \propto l$). But it is straightforward to also assume a more or less linear increase in the value of $V_{eff}$ ($V_{eff} \propto l$) because one does not expect a significant change in the electric field strength and its vertical extension for a fixed gap width. $A/V_{eff}$ would remain approximately constant.
However, the numerical evaluation of $V_{eff}$ for real MM structures shows otherwise. In the following chapter of this publication, we investigate an eSRR MM with a meandering gap structure (ID-eSRR) in simulations and experiments.

Field simulations show that the interdigitated structure exhibits a mode volume of $26~\mu$m$^3$ vs. the $10~\mu$m$^3$ for the eSRR MM with a slot-like gap (cp. Table \ref{tab:Q values}). $V_{eff}$ hence has increased by a factor of 2.6, while the length of the gap and its area are about 8.8 times larger. The ratio $A/V_{eff}$ (again approximating $A$ by the gap width) has increased by a factor of 3.4. According to Equation~(\ref{equ:TNFOM}), this promises an enhanced TN-FOM.

It is worth mentioning that the potential for sensor improvement by lengthening the gap arises from an unexpectedly strong difference between the high-field volume and the mode volume. This difference can be illustrated approximately by calculating how much of the mode volume lies outside the gap area. The results of such calculations are presented in the Supplementary Information, Sec.~C, for the ID-eSRR and eSRR investigated in detail below. For the slot-like gap of the eSRR, we find that—despite the strong field enhancement—most of the mode volume (in our case, more than 98\%) lies outside the gap area. Using a meandering gap brings much more of the mode volume into the gap area (in our case, 11.6\%). This leads to the predicted sensor improvement.

Another parameter in Equation~(\ref{equ:TNFOM}), which can be optimized, is the Q-factor. There are various ways how $Q$ can be improved, e.g., by the use of Fano MMs \cite{Limonov17} or BIC MMs (BIC: Bound states in the continuum) \cite{Abujetas19}. We will not address such options here. We only point out that the use of interdigitated fingers in the gap leads also to a substantial increase in $Q$. Conventional eSRR MMs with a slot-like gap possess a relatively low Q-factor due to radiation loss \cite{Schurig06author}. The introduction of an interdigitated finger structure reduces the radiation loss. The reason is that a significant part of the emission of radiation arises from the gap region. In an interdigitated structure, the dipoles of adjacent minigaps have opposite polarity (as indicated by the black arrows in Figure~\ref{fig:design strategy}(b)), and thus the far-field radiation loss is substantially suppressed. For our structures, the simulations predict that the Q-factor of the ID-eSRR MM is $4\times$ larger than that of the eSRR MM with the slot-like gap.

\section{Interdigitated eSRR (ID-eSRR) metamaterials sensor}
\subsection{Structure design}
\begin{figure}
  \includegraphics[width=0.8\linewidth]{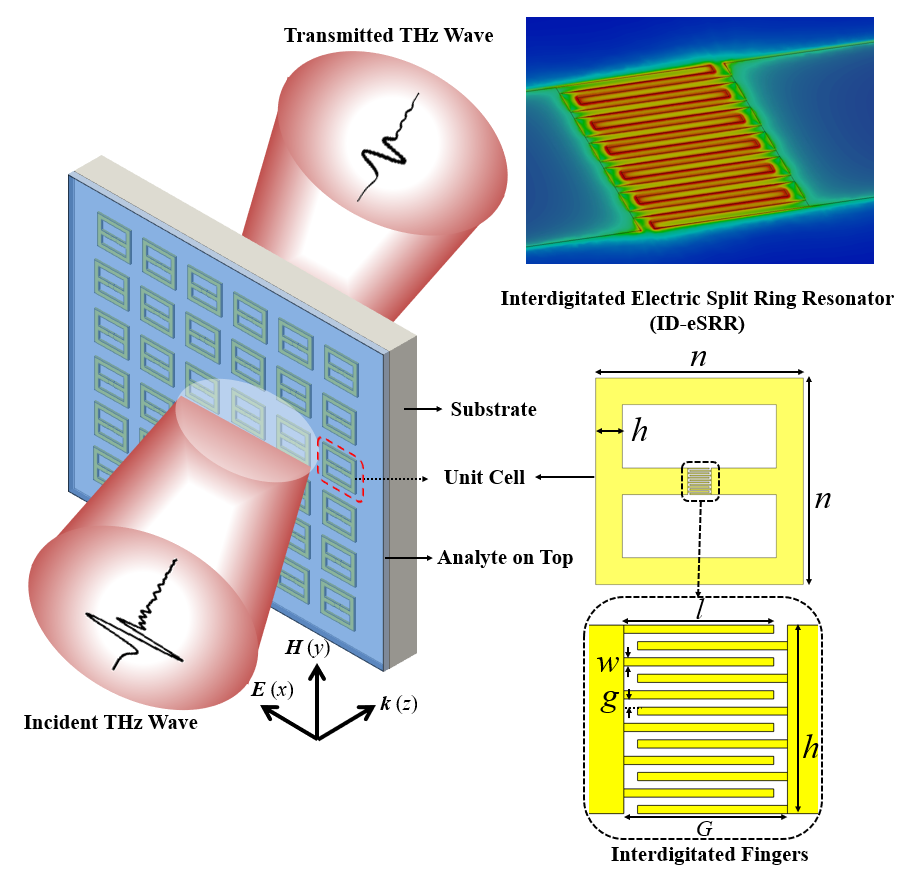}
  \caption{Proposed interdigitated electric split-ring-resonator (ID-eSRR) MMs for THz sensing applications. In the unit cell structure, $n$ is the size of the square ID-eSRR, $h$ the width of the metal stripes, $G$ the gap width, $l$ the finger length, $w$ the finger width, and $g$ the width of the mini-gap between adjacent fingers. 
  }
  \label{fig:structure}
\end{figure}

\textbf{Figure~\ref{fig:structure}} shows the structure of the proposed THz sensor based on an interdigitated electric split-ring resonator (ID-eSRR) on a fused silica substrate. We investigate this structure and compare its properties with those of two traditional eSRR structures, also on fused silica, exhibiting different gap lengths.

The structural parameters of the ID-eSRR and traditional eSRR MMs are listed in \textbf{Table~\ref{tab:parameters}}. They are chosen such that the resonance frequency $f_0$ of the fundamental $LC$ mode of all structures is close to 300~GHz. All devices have square-shaped unit cells with a side length of $p = 240~\mu$m, and the metal stripes forming the SRRs have a width of $h = 13.8~\mu$m in all cases.

Of the two traditional eSSR structures, one has a wide gap with the same extension ($G = 12~\mu$m) as the large gap of the ID-eSRR MM. The other structure has a narrow gap ($G = 0.6~\mu$m) with an extension corresponding to the width of the mini-gap ($g$) of the ID-eSRR structure.

In the ID-eSRR structure with its six pairs of fingers, the finger length ($l = 11~\mu$m) is 1~$\mu$m shorter than the gap width ($G = 12~\mu$m), while the finger width ($w$) and the mini-gap width ($g$) between adjacent fingers remain at a constant value of 0.6~$\mu$m. In terms of the equivalent circuit model for the $LC$ resonance mode, the capacitance ($C$) is primarily determined by the gap region's 
geometry, while the inductance ($L$) mainly arises from the metal stripes (with dimensions $h$, $n$) \cite{Schurig06author, Withayachumnankul10}.

For the ID-eSRR structure, the equivalent capacitance is high due to the large number of finger capacitors in parallel to each other (a total of eleven mini-gaps), with the value of each capacitor determined by the mini-gap width ($g$). Consequently, the equivalent inductance must be lower than that of the traditional eSRR structures to maintain the same resonance frequency. This is achieved by shortening the metal stripes (length $n$).
\\

\begin{table}
\begin{center}
\caption{Structural parameters of the ID-eSRR and eSRR MM sensors working at 300~GHz}
  \begin{tabular}[htbp]{@{}llll@{}}
    \hline
    Parameter ($\mu$m) & ID-eSRR & eSRR ($G$=0.6 $\mu$m) & eSRR ($G$=12 $\mu$m) \\
     \hline
    period ($p$)  & 240 & 240 & 240 \\
    metal stripe length ($n$)  & 107 & 150 & 168 \\
    metal stripe width ($h$)  & 13.8 & 13.8 & 13.8 \\
    gap width ($G$)  & 12 & 0.6 & 12 \\
    finger length ($l$)  & 11 & - & - \\
    finger width ($w$)   & 0.6 & - & -  \\
    mini-gap width ($g$)  & 0.6 & - & -  \\
    \hline
  \end{tabular}
\label{tab:parameters}
\end{center}
\end{table}

\begin{figure}
  \includegraphics[width=\linewidth]{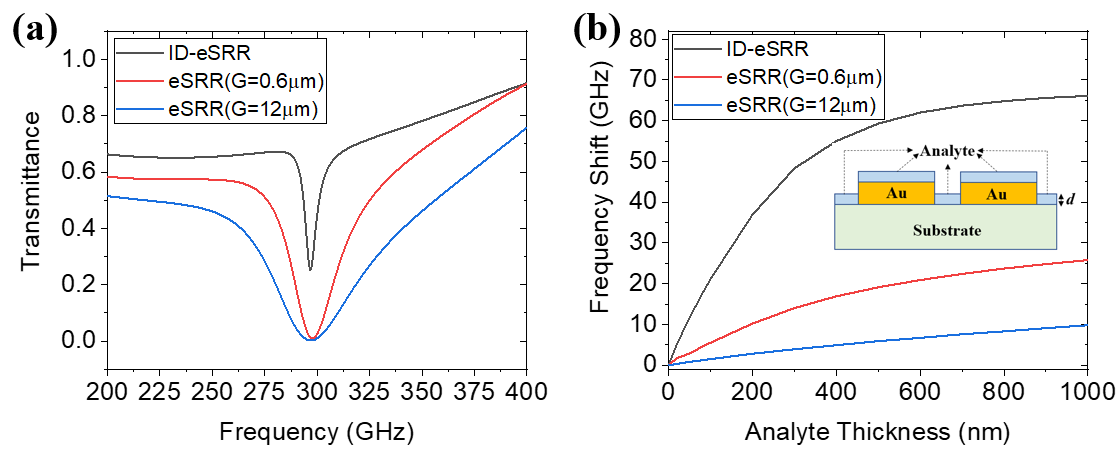}
  \caption{(a) Transmittance spectra of the bare ID-eSRR and eSRR MMs. (b) Refractive index sensing: Frequency shift as a function of analyte thickness ($d$). The inset shows the location of the analyte in all simulations. The analyte material is  $\mathrm{SiO}_2$
 ($\epsilon_r$ = 3.75 + $j \cdot 0.0004$).}
  \label{fig:T_fshift}
\end{figure}

We calculate the transmittance spectra of the three MM structures over the frequency range from 200 to 400~GHz using the CST Studio Suite (Dassault Systèmes).
\textbf{Figure~\ref{fig:T_fshift}}(a) depicts the spectra of the ID-eSRR MM (black line) and the eSRR structures (red and blue lines). The thickness of the gold metallization is assumed to be 200~nm, with a conductivity of gold of $4.56\times10^7$~S/m. The substrate material is fused silica with a thickness of 150~$\mu$m, characterized by a relative permittivity of 3.75 and a loss tangent of 0.0004. The incident THz radiation is polarized along the direction of the interdigitated fingers ($x$-direction in Figure~\ref{fig:structure}).
Comparing the two eSRR structures with each other, one finds that the one with the narrower gap has a sharper resonance than the one with the wider gap. However, the resonance line of the ID-eSRR structure, while not being the deepest, is found to be the sharpest of all three transmittance resonances, indicating that the Q-factor of that structure is higher than those of the two eSRR structures. This feature is beneficial for accurately identifying the resonance frequency during measurements.

\begin{table}
\begin{center}
\caption{Resonance frequency, Q-factor, resonance depth, and mode volume for the ID-eSRR and eSRR MMs}
\begin{tabular}[htbp]{@{}llllll@{}}
\hline
Parameter & ID-eSRR & eSRR ($G=0.6~\mu$m) & eSRR ($G=12~\mu$m) \\
\hline
$f_0$ (GHz)  & 296.3 & 297.5 & 296.9 \\ 
$Q$  & 43.1 & 11.0 & 5.8 \\
$Q_R$  & 113.7 & 12.5 & 6.2\\
$Q_C$  & 71.0 & 89.9 & 99.6 \\
$Q_D$   & 3380.3 & 3122.5 & 1909.6 \\
$T_r$   & 0.42 & 0.56 & 0.49 \\
$V_{eff}$ ($\mu$m$^3$)   & 26 & 10 & 420 \\
\hline
 \end{tabular}
\label{tab:Q values}
\end{center}
\end{table}

\textbf{Table~\ref{tab:Q values}} provides a summary of the properties of the resonance lines of the three structures -- the resonance frequency, the Q-factor and its constituents, and the depth of the resonance. The Q-factor is composed of three contributions which combine to the total Q-factor according to 
\begin{equation}
    \frac{1}{Q} = \frac{1}{Q_R} + \frac{1}{Q_C} + \frac{1}{Q_D}  \,.
 \label{equ:Q factor}
\end{equation}
The three contributions reflect different sources of loss: radiation loss (determining the Q-factor component $Q_R$), conduction or ohmic loss (responsible for $Q_C$), and dielectric loss (leading to $Q_D$). The values of the individual Q-contributions are calculated by numerically varying the material properties of the metal (gold or perfect electrical conductor 'PEC') and the substrate (lossy or loss-free) in the simulations. We then verified that the resultant value of $Q$ is equal to the value obtained by fitting the resonance curve with a Lorentzian function.

For all the structures, the value of $Q_D$ is one or two orders of magnitude higher than $Q_R$ or $Q_C$, indicating that dielectric loss in the substrate is negligible compared to the total losses of the MM structures themselves. Among the three structures, the ID-eSRR exhibits the highest value of $Q$ (43.1), while the eSRR structure with a large gap value ($G = 12~\mu$m) has an extremely low $Q$-value (5.8). For both eSRR structures, radiation loss dominates the total loss, whereas, in the ID-eSRR structure, conduction loss is the dominant factor.

To quantitatively evaluate the radiation loss, a single resonator of ID-eSRR and eSRR with a perfect conductor suspended in air is excited with a current source, and the far-field radiation pattern is captured on a sphere placed one meter away from the resonator. With the introduction of interdigitated fingers, the radiation power density (W/m$^2$) can be reduced by 10~dB from the simulation results of the electric dipole radiation pattern (additional details included in Section A of the Supporting Information). Regarding the resonance depth $T_r$ (absolute value of the difference between the maximum and minimum transmittance around the resonance frequency), the ID-eSRR structure has a lower value compared to the two eSRR structures.\\

\begin{figure}
  \includegraphics[width=\linewidth]{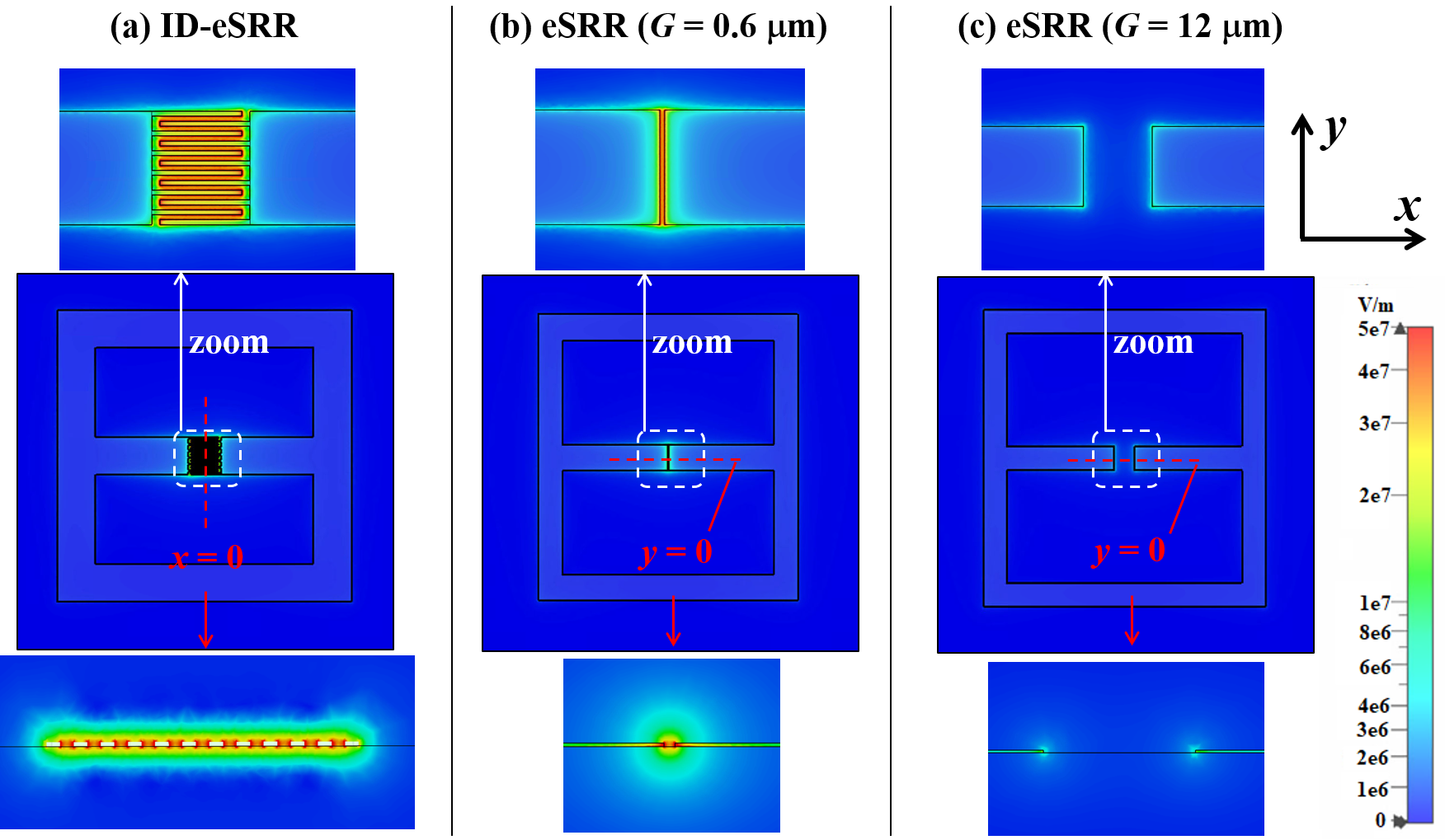}
  \caption{Two top rows of panels: Electric field distribution in the plane of the metal surface at the respective resonance frequency for the three structures (a) ID-eSRR, (b) eSRR ($G = 0.6~\mu$m) and (c) eSRR ($G = 12~\mu$m). Bottom row of pannels: Cuts through the three-dimensional electric field distribution along the red dashed lines shown in the middle row of panels. The amplitude of the incident electric field is $8.09\times10^4$ V/m for all structures.}
  \label{fig:E-field distribution}
\end{figure}

We now compare the mode volumes of the three types of eSRR structures. To determine the mode volumes, we first calculate the field distributions. The six top panels of \textbf{Figure~\ref{fig:E-field distribution}}(a) to (c) display the distribution of the electric field amplitude at the height (200~nm) of the metal surface for the three structures at the respective resonance frequency. In the large-gap eSRR structure (Figure~\ref{fig:E-field distribution}(c)), the strongest electric field is found primarily concentrated near the metal edges of the gap. As the gap width decreases from 12~$\mu$m to 0.6~$\mu$m, the maximum electric field amplitude increases by a factor of five.

The eSRR structure with the narrow gap ($G = 0.6~\mu$m, Figure~\ref{fig:E-field distribution}(b)) exhibits a high electric field covering an area approximately equal to the gap area ($A = h \times G = 8.28~\mu$m$^2$). Similarly in the ID-eSRR structure (Figure~\ref{fig:E-field distribution}(a)), where the field amplitude reaches about the same peak value and the electric field is also concentrated over the entire gap region, which in this case consists of the eleven meandering mini-gaps, covering an area of about $A = 11 \times l \times g = 72.6~ \mu \text{m}^2$. The high-electric-field area hence has increased by 8.8 times by the introduction of the interdigitated fingers. Consequently, this leads to a larger numerator in Equation~(\ref{equ:simplified perturbation}), and significantly larger frequency shifts can be expected upon analyte sensing.\\

Before continuing with the mode volume, we point out a feature of the field distribution that is important for sensing applications. The three panels at the bottom of Figure~\ref{fig:E-field distribution} display two-dimensional cuts through the three-dimensional electric fields along the red dashed lines shown in the middle row of panels. There is a remarkable feature in Figs.~\ref{fig:E-field distribution}(a) and (b): The electric field in the gaps is mainly concentrated in the air (or the vacuum) between the metal edges, and only weakly extends into the substrate. This feature is attributed to the reduction of the fringing field as the distance between the metal edges (0.6~$\mu$m) decreases to a similar scale as the height of the metal edges (0.2~$\mu$m). The benefit of this field concentration is that it is in the free space of the gaps and is available for interaction with an analyte.

We now turn to the mode volume of the three types of structures. Table~\ref{tab:Q values} lists the values of the mode volume calculated with Equation~(\ref{equ:mode volume}) for the three structures. The mode volume of the small-gap eSRR structure ($10~\mu$m$^3$) is only 2.4\% of that of the large-gap eSRR structure ($420~\mu$m$^3$). By introducing interdigitated fingers, the mode volume of the ID-eSRR structure is raised to $26~\mu$m$^3$, which amounts to 2.6 times the mode volume of the eSRR structure with $G = 0.6~\mu$m. Expressed in terms of the resonance wavelength $\lambda_0$ in free space, it is equal to $2.6 \times 10^{-9} \times \lambda_0^3$.

It is apparent that the mode volume does not increase as rapidly as the area $A$. The reason is that the electric energy contained within the gaps is only a small portion of the total energy stored in the resonator. For example, this factor is only 11.6\% for the ID-eSRR structure and 1.76\% for the eSRR structure with $G$=0.6~$\mu$m. Increasing the high-electric-field area does not proportionally increase the total energy. (Detailed information regarding this factor is provided in Section C of the Supporting Information). From those numbers and according to Equation~(\ref{equ:simplified perturbation}), the sensitivity and the TN-FOM of the ID-eSRR structure are estimated to increase by about 3.5 and 15 times, respectively, than that of the small-gap eSRR.\\

\subsection{Simulated sensing performances}

The resonance frequency experiences a redshift when a homogeneous layer of dielectric analyte is deposited onto the MM. The extent of frequency shift depends on both the thickness and refractive index of the analyte. We perform CST simulations for an analyte for which we assume its complex-valued relative permittivity to be $3.75 + j \cdot 0.0004$ (equal to that of the fused silica substrate). We assume in the simulations that the analyte is an interrupted layer, as shown in the inset of Figure~\ref{fig:T_fshift}(b). Figure~\ref{fig:T_fshift}(b) compares the resonance frequency shift of the ID-eSRR and eSRR MMs as a function of the thickness of the analyte layer (20~nm to 1000~nm). The resonance frequency shift exhibits a nonlinear relationship with the analyte thickness due to the exponential decay of the evanescent electric field from the metal surface to the upper air region \cite{Cao22}. Therefore, as the analyte thickness increases, the frequency shift initially rises rapidly, then increases more gradually before eventually saturating. When the analyte thickness is 100~nm, the frequency shift for the ID-eSRR structure is 21.1~GHz, which is 14.1 times that of the large-gap eSRR structure and 3.8 times that of the eSRR structure with the small gap. With an analyte thickness of 1000~nm, the frequency shift for the ID-eSRR structure increases to 66.1 GHz, which is 2.6 times (respectively 6.7 times) the value for the large-gap (small-gap) eSRR structure. The ID-eSRR MM sensor promises hence to be much more sensitive than the eSRR MM sensors, especially so if the analyte layer is thin.\\

\begin{table}[h]
\begin{center}
 \caption{Sensing performances of the ID-eSRR and eSRR MMs}
 \begin{tabular}[htbp]{@{}lllll@{}}
\hline
Parameter & Analyte thickness (nm) & ID-eSRR & eSRR ($G = 0.6~\mu$m) & eSRR ($G = 12~\mu$m) \\
\hline
\multirow{3}{*}{$S$ (GHz/RIU)} & 100  & 22.6 & 5.9 & 1.6 \\ 
~ & 500  & 63.0 & 20.5 & 6.3 \\
~ & 1000  & 70.1 & 27.7 & 10.5\\
\hline
\multirow{3}{*}{FOM (1/RIU)} & 100  & 3.29 & 0.22 & 0.03 \\
~ & 500 & 9.16 & 0.76 & 0.12 \\
~ & 1000 & 10.20 & 1.02 & 0.21 \\
\hline
\multirow{3}{*}{TN-FOM (1/($\mu$m$\cdot \text{RIU}$))} & 100  & 32.87 & 2.18 & 0.31 \\
~ & 500 & 18.33 & 1.52 & 0.25 \\
~ & 1000 & 10.20 & 1.02 & 0.21 \\
\hline
\end{tabular}
\label{tab:Sensitivity}
\end{center}
\end{table}

\textbf{Table~\ref{tab:Sensitivity}} compares the calculated values of the sensitivity $S$ (in GHz/RIU), the figure of merit (FOM, in units of RIU$^{-1}$), and the thickness normalized FOM (TN-FOM, in units of 1/($\mu$m$\cdot \text{RIU}$)) of the three sensors. The ID-eSRR MMs exhibit advantages in terms of high values of the Q-factor, sensitivity, and FOM. When the analyte thickness is 100~nm, the sensitivity of the ID-eSRR structure is 3.8 times, and the FOM (TN-FOM) value is 15.1 times that of the small-gap eSRR MM. The numbers from these independent simulations are consistent with the estimations from Equation~(\ref{equ:simplified perturbation}), as mentioned in the previous section, confirming the validity of perturbation theory. Putting this performance into perspective with the literature on MM sensors, we note that the ID-eSRR MM has a higher sensitivity for a layer thickness of 100~nm than the device of Ref.~\cite{Withayachumnankul12} reaches for an analyte thickness of 2.17 $\mu$m (8.1~GHz/RIU). It should be noted that the frequency shift and sensitivity of the ID-eSRR structure can be further enhanced by decreasing the width of the mini-gaps (g). Additional information regarding this improvement is provided in Section D of the Supporting Information.\\

\subsection{Fabrication details}
\begin{figure}[h]
  \includegraphics[width=\linewidth]{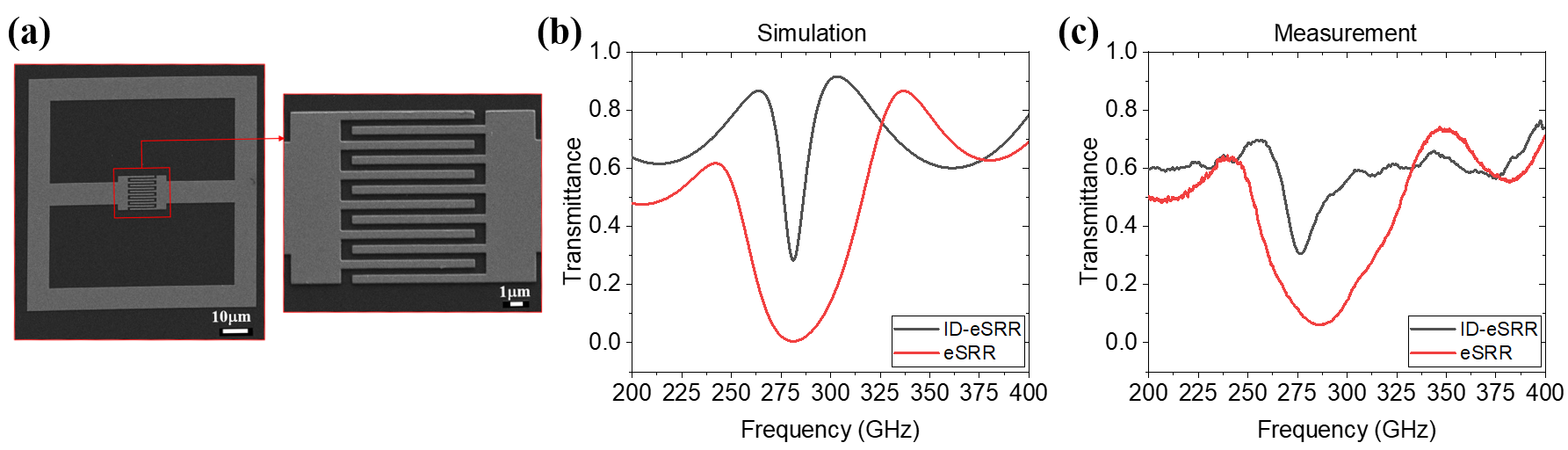}
  \caption{(a) SEM image of an ID-eSRR structure fabricated by electron beam lithography, (b) simulated transmittance spectra of both the ID-eSRR MM ($G = 12~\mu$m, $w = g= 0.6~\mu$m) and the eSRR MM ($G = 12~\mu$m) without analyte, (c) measured transmittance spectra of the ID-eSRR and eSRR MMs without analyte.}
  \label{fig:Measurement}
\end{figure}

To validate the simulation results, we fabricated two MMs via electron beam lithography (EBL). One of them features an ID-eSRR design with parameters $G = 12~\mu$m and $w = g = 0.6~\mu$m. The other structure serves as a reference; it has an eSRR configuration and a gap without fingers with $G = 12~\mu$m. The ID-eSRR MM consists of 18 $\times$ 18 unit cells, while the eSRR MM has 12 $\times$ 12 unit cells, and the two MMs have a size of 3.076 $\times$ 3.076~mm$^2$. The substrate is 500~$\mu$m thick.
The fabrication process started with the deposition of a thick layer of polymethyl methacrylate (PMMA) 950A11 (1.8 ~$\mu$m) on a fused silica substrate with a size of 40 $\times$ 40 ~mm$^2$. Due to the transparent nature of the fused silica, an additional photoconductive polymer layer (40~nm) was deposited over the PMMA layer (AR-PC 5092.02). Two EBL exposure steps were performed to design the MMs. The first exposure was used to create the markers used to detect the sample height in the EBL machine. The second exposure was performed to create the MMs with different features. After each exposure, a metallization process and a lift-off process were performed. The 10/200 nm thick Cr/Au films were deposited by e-beam evaporation after the second EBL exposure. The scanning-electron-microscope (SEM) image of a single ID-eSRR resonator is shown in \textbf{Figure~\ref{fig:Measurement}}(a).

\section{Experimental results}
Transmission THz measurements are performed with a continuous-wave THz spectroscopy system operating in the frequency domain. This system is based on a TeraScan 1550 system from TOPTICA Photonics AG, which offers high bandwidth, a high dynamic range, and a spectral resolution of 1~MHz \cite{Anselm15}. To adapt it for the characterization of the MM sensors, four off-axis parabolic mirrors (focal length of 4 inches and diameter of 2 inches) are incorporated to create a quasi-optical transmission path with an intermediate focal plane, where the MM sensor is positioned. A vacuum holder for the specimens is mounted on motorized linear stages (MICOS LS-65). It is designed to enable reproducible, precise mounting of the MMs. During all measurements, spectral scans are performed with a scan speed of 240~GHz/min and a lock-in integration time of 10~ms. All scans are repeated five times, and the mean value is taken for the evaluation and plotting of the transmission spectra.

In Figures~\ref{fig:Measurement}(b) and (c), we present the simulated and measured transmittance spectra, respectively, for both types of MM sensors. The agreement between the resonance frequencies, the amplitudes, and the shapes of the simulated and measured spectra is quite satisfactory. 
The remaining discrepancies may stem from the different illumination conditions in the simulations and measurements: The simulations assume a plane wave with a well-defined wave-vector, while the measurements are performed with a focused THz beam with an angular spread of wave-vectors, for which the MM's response may vary (further details are provided in Section E of the Supporting Information). Additional deviations may arise from different values of the substrate permittivity and the metal conductivity in the experiment and simulations, as well as by fabrication tolerances, particularly with regard to the metal fingers.

\begin{figure}[ht]
  \includegraphics[width=\linewidth]{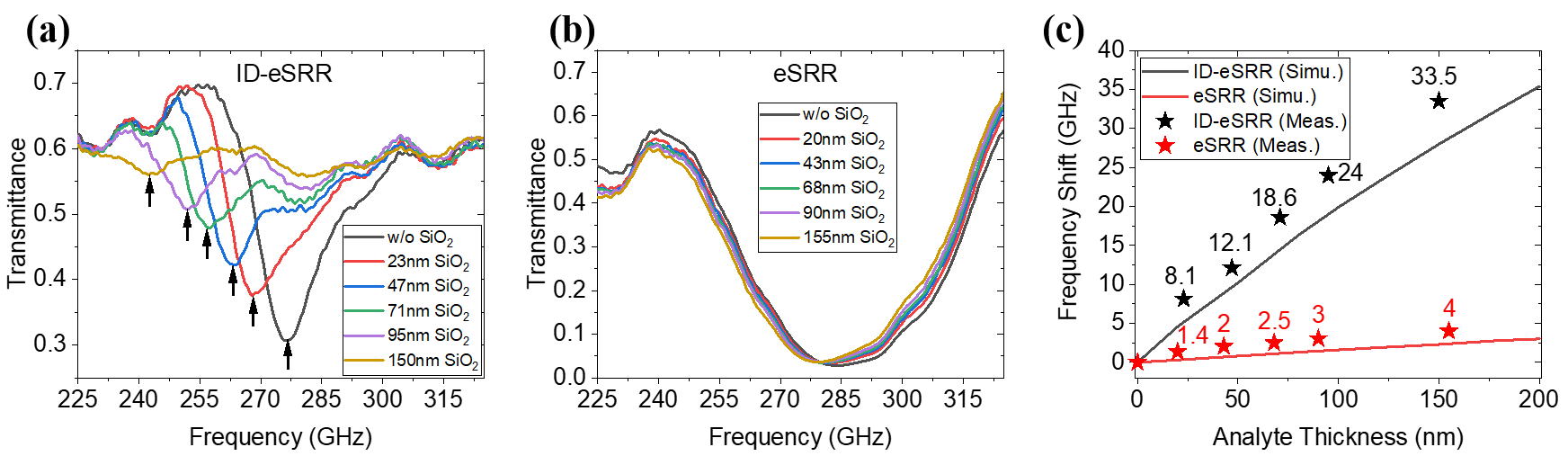}
  \caption{(a) Measured transmittance spectra of the ID-eSRR structure loaded with a $\mathrm{SiO}_2$ layer of varying thickness (from 23 to 150~nm). (b) Likewise, but for the eSRR structure ($\mathrm{SiO}_2$ layer thickness varying from 20 to 155~nm). (c) Simulated and measured resonance frequency shift in dependence of the $\mathrm{SiO}_2$ layer  thickness.}
  \label{fig:Freqshift_meas}
\end{figure}

In order to assess the sensitivity of the sensors, we deposit layers of $\mathrm{SiO}_2$ with varying thickness onto the sensor surface using a self-built RF magnetron sputtering chamber with a $\mathrm{SiO}_2$ target, offering a sputter rate of around 0.25~nm/s. We begin with a thin layer, then measure the transmittance, and then deposit more $\mathrm{SiO}_2$ onto the specimen for the next measurement cycle. This process is repeated several times. 
During each sequence, the actual thickness of the $\mathrm{SiO}_2$ layer is determined with a surface profilometer (Bruker Nano-Dektak). Before each sputtering process as well as before each THz measurement, the MM sensor is cleaned with acetone (3~min) and isopropanol (3~min), and then dried in nitrogen gas flow to ensure that the surface of the sensor is free from contamination.

\textbf{Figure~\ref{fig:Freqshift_meas}}(a,b) presents the measured transmittance spectra for the empty and analyte-loaded ID-eSRR and eSRR structures. Data for five distinct values of the $\mathrm{SiO}_2$ layer thickness are shown. These are for the ID-eSRR sensor 23~nm, 47~nm, 71~nm, 95~nm and 150~nm, and for the eSRR sensor 20~nm, 43~nm, 68~nm, 90~nm and 155~nm. For each spectrum, we determine the respective resonance frequency. In the case of the ID-eSRR sensor, this is simply the frequency position of the transmission minimum (marked by an upwards-pointing black arrow in Figure~\ref{fig:Freqshift_meas}(a)). For the  eSRR sensor, the minimum is difficult to identify directly. We determine it by fitting the resonance curve with a Lorentzian function around the broad transmission valley. It is obvious that the presence of the analyte leads to a much stronger shift of the resonance frequency in the case of the ID-eSRR MM as compared with the eSRR MM. It should be noted that the resonance depth of the ID-eSRR sensor decreases with the increase of analyte thickness. This modulation is mainly caused by the Fabry-Perot resonance within the thick substrate. Further details are provided in Section F of the Supporting Information.

Figure~\ref{fig:Freqshift_meas}(c) displays the resultant frequency shifts for the two structures. For comparison, we also include the simulated frequency shifts. Both theory and experiment confirm the superior sensitivity of the ID-eSRR structure.
At a thickness of around 20~nm, the ID-eSRR structure exhibits a measured frequency shift of 8.1~GHz, compared to a value of 1.4~GHz for the eSRR structure, a nearly six times better performance. For a thickness of about 150~nm, this ratio is even larger, reaching a value of 8.4. There, the Q-factor, sensitivity, and FOM value of the ID-eSRR structure are higher by factors of 6.6, 8.4, and 55.3, respectively. 
For both structures, a weak trend towards saturation is observed.

\begin{table}
\begin{center}
 \caption{Performance comparison with other THz metamaterial sensors described in the literature. Data derived from experiments.}
   \begin{threeparttable}
  \begin{tabular}[htbp]{@{}llllllll@{}}
    \hline
    \thead {Analyte \\ thickness}  & \thead {$f_0$ \\ (GHz)} & $Q$ & \thead {$S$ \\ (GHz/RIU)}  & \thead {FOM \\ (1/RIU)} & \thead {TN-FOM \\ (1/($\mu$m$\cdot \text{RIU}$)} & \thead {Unit cell \\ structure} & Reference \\
     \hline
    2.17 $\mu$m  & 407 & 4 & 8.1 & 0.08 & 0.04 & SRR & \cite{Withayachumnankul12} \\
    1 $\mu$m  & 515 & 28\tnote{*} & 16.7 & 0.91\tnote{*} & 0.91\tnote{*} & aDSRR & \cite{Singh14} \\
    250 nm  & 422.6 & 9.6\tnote{*} & 6.0 & 0.14\tnote{*} & 0.54\tnote{*} & Toroidal aDSRR & \cite{Gupta17} \\
    240 nm  & 600 & 10\tnote{*} & 18.0 & 0.30\tnote{*} & 1.25\tnote{*} & I-shaped structure & \cite{Gupta20}\\
    40 nm  & 990 & 20 & 10.8 & 0.22 & 5.45 & aDSRR & \cite{Srivastava19} \\
    47 nm  & 276 & 17.7 & 12.1 & 0.78 & 16.51 & ID-eSRR & this work\\
   150 nm  & 276 & 17.7 & 33.5 & 2.15 & 14.32 & ID-eSRR & this work\\
    \hline
  \end{tabular}
      \begin{tablenotes}
      \item[*] Simulation values are used because the experimental values are not available.
    \end{tablenotes}
    \end{threeparttable}
\label{tab:senscomp}
\end{center}
\end{table}

\section{Discussion}
One notices in Figure~\ref{fig:Freqshift_meas}(c) that the measured frequency shift is always larger than the calculated one. We attribute this systematic difference to the treatment of the analyte layer in the simulations as an interrupted film (Figure~\ref{fig:T_fshift}(b)). Because the film thickness is inferior to the MM thickness (200~nm), there always exists additional analyte on vertical edges (the $yoz$ plane) of fingers in the sputtering process, which cannot be properly considered in the simulation. As the electric field is strongly concentrated in that gap, the simulations underestimate the strength of the interaction between the field and the analyte material.

We now come to a comparison of the \textit{measured} performance parameters of the ID-eSRR MM sensor with the literature. In \textbf{Table~\ref{tab:senscomp}}, we list analyte thicknesses, resonance frequencies, Q-factors, $S$ and FOM (TN-FOM) values as well as the unit cell structures of various MM sensors. Where experimental data are not available (see values in Table~\ref{tab:senscomp} marked by an asterisk), we list simulation results taken from the publications. When making comparisons, one should bear in mind that the respective literature studies have been performed not only at different resonance frequencies but also for different types of analytes and values of the analyte thickness. As already demonstrated by the simulation results for our sensors in Table~\ref{tab:Sensitivity}, both $S$ and FOM (which correct for the different refractive-index values of the various analytes) increase with analyte thickness; a feature that is also true for the sensors studied in the literature.

However, Table~\ref{tab:senscomp} still allows us to make some statements on the relative performance of the ID-eSRR MM. At an analyte thickness of 150~nm, the FOM value of the ID-eSRR structure outperforms that of all other reported sensors, most of which were subjected to the tests with thicker layers. Only for one MM, the literature reports measurements at small thickness; this is the aDSRR sensor of Ref.~\cite{Srivastava19}, a state-of-the-art device whose operation principle is based on a Fano resonance. To keep the field concentrated above the substrate, it was built on a membrane with a low dielectric constant; the resonance frequency was chosen high (990~GHz) for a high field extension into the air. Even so, it reaches, at an analyte thickness of 40~nm, a value of $S$ that is quite similar to that obtained with our ID-eSSR MM at a thickness of 47~nm; however, the FOM and TN-FOM achieved with the ID-eSSR sensor is considerably higher.

Finally, we compare with the performance parameters of the I-shaped sensor of Ref.~\cite{Gupta20}, which was also operated at a higher resonance frequency (approximately 600~GHz) than our ID-eSRR device. For a $\mathrm{SiO}_2$ layer thickness of 150~nm, the latter reaches better values of $S$ and FOM than were obtained with the I-shaped sensor at a $\mathrm{SiO}_2$ layer thickness of 240~nm. From Figure~\ref{fig:Freqshift_meas}(a,c) and Ref.~\cite{Gupta20}, we find that the ID-eSRR MM sensor reaches about the same absolute frequency shift (18.6~GHz) as the I-shaped sensor but already for an $\mathrm{SiO}_2$ layer thickness of 71~nm instead of 240~nm, and this at a much lower resonance frequency.

These results corroborate the exceptional sensitivity of the proposed ID-eSRR sensor in detecting analytes and that it is especially well-suited for low analyte thickness. 

\section{Conclusion}
In conclusion, we have proposed a general optimization method for designing high-performance planar metamaterial sensors operating at terahertz frequencies. This method relies on simplified dielectric perturbation theory. By utilizing resonators with high Q-factors, reduced mode volumes, and enhancing the overlap volume between the high-electric-field region and the analyte material, sensor sensitivity can be significantly improved. This approach serves as a versatile method for the sensitive detection of thin dielectric films and minute traces of molecules.

As a demonstration and application of dielectric perturbation theory, we designed a sensitive metamaterial-based terahertz sensor for thin-film analytes. The metamaterial's unit cell comprises an electric split-ring resonator with a gap in the metal stripes implemented as an interdigitated finger structure. The device is tailored for a resonance frequency of 0.3 THz. In comparison with its parent metamaterial featuring a slot-like gap (denoted as length $G$ in the main text) instead of the meander-like gap between interdigitated fingers, the novel structure exhibits a significantly higher Q-factor and strong electric-field enhancement over a sizable area. Consequently, this leads to a significantly enhanced sensitivity, figure of merit (FOM), and thickness-normalized FOM. In a proof-of-principle experiment involving a SiO$_2$ analyte layer, we observed a more than fiftyfold increase in FOM (TN-FOM).

This innovative sensor holds promise for label-free and amplification-free detection of trace amounts of analytes, particularly biomolecules such as DNA and proteins. Importantly, the sensors can be implemented across a wide range of frequencies due to the frequency-scalable nature of the design.
\\


\medskip
\textbf{Supporting Information} \par 
Supporting Information is available from the Wiley Online Library or from the first author.

\textbf{Acknowledgements} \par 
This research work was funded by DFG projects RO 770/46-1, RO 770/50-1 and HA3022/15 (the latter two being part of the  DFG-Schwerpunkt ``Integrierte Terahertz-Systeme mit neuartiger Funktionalität'' (INTEREST -- SPP 2314)). Lei Cao acknowledges support from the HUST Overseas Training Program for Outstanding Young Teachers. The computation is completed in the HPC Platform of Huazhong University of Science and Technology.

\textbf{Conflict of Interest} \par 
The authors declare no conflict of interest.

\medskip

%
\bibliography{Metamaterials}

\begin{thebibliography}{10}
\providecommand{\url}[1]{\texttt{#1}}
\providecommand{\urlprefix}{URL }

\bibitem{Withayachumnankul09}
W.~Withayachumnankul, D.~Abbott,
\newblock \emph{IEEE Photonics Journal} \textbf{2009}, \emph{1}, 2 99.

\bibitem{Debus07}
C.~Debus, P.~H. Bolivar,
\newblock \emph{Applied Physics Letters} \textbf{2007}, \emph{91}, 18.

\bibitem{Al-Naib08}
I.~A.~I. Al-Naib, C.~Jansen, M.~Koch,
\newblock \emph{Applied Physics Letters} \textbf{2008}, \emph{93}, 8 083507.

\bibitem{Singh14}
R.~Singh, W.~Cao, I.~Al-Naib, L.~Cong, W.~Withayachumnankul, W.~Zhang,
\newblock \emph{Applied Physics Letters} \textbf{2014}, \emph{105}, 17.

\bibitem{Weisenstein20}
C.~Weisenstein, D.~Schaar, A.~K. Wigger, H.~Schafer-Eberwein, A.~K. Bosserhoff, P.~H. Bolivar,
\newblock \emph{Biomedical Optics Express} \textbf{2020}, \emph{11}, 1 448.

\bibitem{Park17}
S.~J. Park, S.~H. Cha, G.~A. Shin, Y.~H. Ahn,
\newblock \emph{Biomedical Optics Express} \textbf{2017}, \emph{8}, 8 3551.

\bibitem{Meng22}
F.~Meng, F.~Han, U.~Kentsch, A.~Pashkin, C.~Fowley, L.~Rebohle, M.~D. Thomson, S.~Suzuki, M.~Asada, H.~G. Roskos,
\newblock \emph{Optics Letters} \textbf{2022}, \emph{47} 4969.

\bibitem{Withayachumnankul12}
W.~Withayachumnankul, H.~Lin, K.~Serita, C.~M. Shah, S.~Sriram, M.~Bhaskaran, M.~Tonouchi, C.~Fumeaux, D.~Abbott,
\newblock \emph{Optics Express} \textbf{2012}, \emph{20}, 3 3345.

\bibitem{Islam17}
M.~Islam, S.~J.~M. Rao, G.~Kumar, B.~P. Pal, D.~R. Chowdhury,
\newblock \emph{Scientific Reports} \textbf{2017}, \emph{7}, 1 7355.

\bibitem{Jun22}
S.~W. Jun, Y.~H. Ahn,
\newblock \emph{Nat Commun} \textbf{2022}, \emph{13}, 1 3470.

\bibitem{Gupta20}
M.~Gupta, R.~Singh,
\newblock \emph{Advanced Optical Materials} \textbf{2020}, \emph{8}, 16 1902025.

\bibitem{Hu16}
X.~Hu, G.~Xu, L.~Wen, H.~Wang, Y.~Zhao, Y.~Zhang, D.~R.~S. Cumming, Q.~Chen,
\newblock \emph{Laser \& Photonics Reviews} \textbf{2016}, \emph{10}, 6 962.

\bibitem{Jauregui18}
I.~Jáuregui-López, P.~Rodríguez-Ulibarri, A.~Urrutia, S.~A. Kuznetsov, M.~Beruete,
\newblock \emph{physica status solidi (RRL) - Rapid Research Letters} \textbf{2018}, \emph{12}, 10.

\bibitem{Saadeldin19}
A.~S. Saadeldin, M.~F.~O. Hameed, E.~M.~A. Elkaramany, S.~S.~A. Obayya,
\newblock \emph{IEEE Sensors Journal} \textbf{2019}, \emph{19}, 18 7993.

\bibitem{Driscoll07}
T.~Driscoll, G.~O. Andreev, D.~N. Basov, S.~Palit, S.~Y. Cho, N.~M. Jokerst, D.~R. Smith,
\newblock \emph{Applied Physics Letters} \textbf{2007}, \emph{91}, 6 062511.

\bibitem{Al-Naib17}
I.~Al-Naib,
\newblock \emph{IEEE Journal of Selected Topics in Quantum Electronics} \textbf{2017}, \emph{23}, 4 1.

\bibitem{Jin18}
B.~Jin, W.~Tan, C.~Zhang, J.~Wu, J.~Chen, S.~Zhang, P.~Wu,
\newblock \emph{Advanced Theory and Simulations} \textbf{2018}, \emph{1}, 9.

\bibitem{Ma20author}
J.~Ma, S.~Wang, Y.~Yang, K.~Wang, L.~Guo, Y.~Gong,
\newblock \emph{AIP Advances} \textbf{2020}, \emph{10}, 8.

\bibitem{Sun22}
L.~Sun, L.~Xu, J.~Y. Wang, Y.~N. Jiao, Z.~H. Ma, Z.~F. Ma, C.~Chang, X.~Yang, R.~D. Wang,
\newblock \emph{Nanoscale} \textbf{2022}, \emph{14}, 27 9681.

\bibitem{Meng23}
F.~Meng, L.~Cao, A.~Karalis, H.~Gu, M.~D. Thomson, H.~G. Roskos,
\newblock \emph{Optics Express} \textbf{2023}, \emph{31}, 24.

\bibitem{Tang20}
M.~Tang, L.~Xia, D.~Wei, S.~Yan, M.~Zhang, Z.~Yang, H.~Wang, C.~Du, H.~L. Cui,
\newblock \emph{Spectrochimica Acta Part A: Molecular and Biomolecular Spectroscopy} \textbf{2020}, \emph{228} 117736.

\bibitem{Gezimati23}
M.~Gezimati, G.~Singh,
\newblock \emph{Optical and Quantum Electronics} \textbf{2023}, \emph{55}, 8.

\bibitem{Weisenstein21}
C.~Weisenstein, A.~K. Wigger, M.~Richter, R.~Sczech, A.~K. Bosserhoff, P.~H. Bolivar,
\newblock \emph{Journal of Infrared, Millimeter, and Terhertz Waves} \textbf{2021}, \emph{42} 607.

\bibitem{Richter22}
M.~Richter, Y.~Loth, C.~Weisenstein, A.~K. Wigger, D.~Schaar, A.~K. Bosserhoff, P.~H. Bolívar,
\newblock \emph{Frequenz} \textbf{2022}, \emph{76}, 11-12 627.

\bibitem{Askari21author}
M.~Askari, H.~Pakarzadeh, F.~Shokrgozar,
\newblock \emph{Journal of the Optical Society of America B} \textbf{2021}, \emph{38}, 12 3929.

\bibitem{Pozar11}
D.~Pozar,
\newblock \emph{Microwave Engineering, 4th Edition},
\newblock Wiley, \textbf{2011}.

\bibitem{Cao22}
L.~Cao, S.~S. Jia, M.~D. Thomson, F.~Q. Meng, H.~G. Roskos,
\newblock \emph{Optics Express} \textbf{2022}, \emph{30}, 8 13659.

\bibitem{Seo09}
M.~A. Seo, H.~R. Park, S.~M. Koo, D.~J. Park, J.~H. Kang, O.~K. Suwal, S.~S. Choi, P.~C.~M. Planken, G.~S. Park, N.~K. Park, Q.~H. Park, D.~S. Kim,
\newblock \emph{Nature Photonics} \textbf{2009}, \emph{3}, 3 152.

\bibitem{Bahk17}
Y.-M. Bahk, S.~Han, J.~Rhie, J.~Park, H.~Jeon, N.~Park, D.-S. Kim,
\newblock \emph{Physical Review B} \textbf{2017}, \emph{95}, 7 075424.

\bibitem{Adak19}
S.~Adak, L.~N. Tripathi,
\newblock \emph{Analyst} \textbf{2019}, \emph{144}, 21 6172.

\bibitem{Gerry04}
C.~Gerry, P.~Knight,
\newblock \emph{Introductory Quantum Optics},
\newblock Cambridge University Press, Cambridge, \textbf{2004}.

\bibitem{Yan19}
X.~Yan, M.~S. Yang, Z.~Zhang, L.~J. Liang, D.~Q. Wei, M.~Wang, M.~J. Zhang, T.~Wang, L.~H. Liu, J.~H. Xie, J.~Q. Yao,
\newblock \emph{Biosensors \& Bioelectronics} \textbf{2019}, \emph{126} 485.

\bibitem{Abdolrazzaghi18}
M.~Abdolrazzaghi, M.~Daneshmand, A.~K. Iyer,
\newblock \emph{IEEE Transactions on Microwave Theory and Techniques} \textbf{2018}, \emph{66}, 4 1843.

\bibitem{CaoY22}
Y.~Cao, C.~Ruan, K.~Chen, X.~Zhang,
\newblock \emph{Scientific Reports} \textbf{2022}, \emph{12}, 1 1255.

\bibitem{Srivastava19}
Y.~K. Srivastava, R.~T. Ako, M.~Gupta, M.~Bhaskaran, S.~Sriram, R.~Singh,
\newblock \emph{Applied Physics Letters} \textbf{2019}, \emph{115}, 15.

\bibitem{Limonov17}
M.~F. Limonov, M.~V. Rybin, A.~N. Poddubny, Y.~S. Kivshar,
\newblock \emph{Nature Photonics} \textbf{2017}, \emph{11}, 9 543.

\bibitem{Abujetas19}
D.~R. Abujetas, N.~van Hoof, S.~ter Huurne, J.~G. Rivas, J.~A. Sánchez-Gil,
\newblock \emph{Optica} \textbf{2019}, \emph{6}, 8.

\bibitem{Schurig06author}
D.~Schurig, J.~J. Mock, D.~R. Smith,
\newblock \emph{Applied Physics Letters} \textbf{2006}, \emph{88}, 4 041109.

\bibitem{Withayachumnankul10}
W.~Withayachumnankul, C.~Fumeaux, D.~Abbott,
\newblock \emph{Optics Express} \textbf{2010}, \emph{18}, 25 25912.

\bibitem{Anselm15}
A.~J. Deninger, A.~Roggenbuck, S.~Schindler, S.~Preu,
\newblock \emph{Journal of Infrared, Millimeter, and Terahertz Waves} \textbf{2015}, \emph{36}, 3 269.

\bibitem{Gupta17}
M.~Gupta, Y.~K. Srivastava, M.~Manjappa, R.~Singh,
\newblock \emph{Applied Physics Letters} \textbf{2017}, \emph{110}, 12 121108.

\end{thebibliography}
\bibliographystyle{MSP}





\end{document}